\documentstyle[12pt]{article}
\def\beq{\begin{equation}}
\def\eeq{\end{equation}}
\def\bey{\begin{eqnarray}}
\def\eey{\end{eqnarray}}
\def\b{\bibitem}
\begin{document}
\baselineskip .22in
\begin{flushright}
PUPT-1310\\
March 1992\\
\end{flushright}
\vspace{20mm}
\begin{center}
THE WEINGARTEN MODEL\\
\`{A} LA POLYAKOV\\
\vspace{20mm}
{\bf Simon Dalley}\\
{\em Joseph Henry Laboratories, Department of Physics,\\
Princeton University, Princeton N.J.08544, U.S.A.}\\
\vspace{20mm}
{\bf Abstract}\\
\end{center}

The Weingarten lattice gauge model of Nambu-Goto strings is generalised to
allow for fluctuations of an intrinsic worldsheet metric through a dynamical
quadrilation. The continuum limit is taken for $c\leq 1$ matter, reproducing
the results of hermitian matrix models to all orders in the genus expansion.
For the compact $c=1$ case the vortices are Wilson lines, whose exclusion
leads to a theory of non-interacting fermions. As a by-product
of the analysis one
finds the critical behaviour of SOS and vertex models coupled to 2D quantum
gravity.
\vfil
\newpage
\section{Introduction}
The motivation for the work presented in this note
 was to make the old analogy between string and
(lattice) gauge theories an equivalence. This will be achieved through a
modification \`{a} la Polyakov of Weingarten's generalisation of lattice
Yang-Mills theory \cite{wein}. Although the cut-off
bosonic theory will be generally defined, the usual problems of finding a
continuum limit for $c>1$ matter will mean that quantitative results can
only be obtained for $c\leq1$. In fact work on this latter problem was
initiated some time ago by T.R.Morris \cite{timx}.
In this letter it is proved that there are continuum
limits on surfaces of arbitrary genus equivalent to those of the multi-critical
hermitian matrix (open) chain models \cite{shen,doug,gros}.
The main reason
for introducing a lattice gauge formulation however appears in the finite
temperature theory, described by compact $c=1$. In the new variables vortices
are manifest, since they are just Wilson lines, and their exclusion leads
to a simple proof of equivalence to non-interacting fermions at finite
temperature (the singlet sector of hermitian matrix models).
The phase transition due to a condensation of winding modes can
now be couched in the lore of `deconfinement'. Dual to this phenomenon is a
condensation of Kaluza-Klein momentum modes if the radius, for fixed
number of target lattice sites, is made too
large; the theory is equivalent to that on a continuous target circle
provided the lattice spacing is less than its critical value \cite{igor3}.
As a subsidiary result, the analyses exhibit the critical
behaviour of certain face and vertex models on random surfaces. In the
discussion at the end, some speculations and suggestions for more detailed
investigation are given.
\section{The Weingarten Model}
The original Weingarten
model of Nambu-Goto strings was constructed as follows. On a
$D$-dimensional hypercubic lattice with oriented links $\pm l$ one associates
a complex $N$x$N$ matrix $M(l)$ ($M(-l)=M^{\dagger}(l)$) to each link $l$.
The partition function is;
\beq
Z_{W} = \int_{-\infty}^{+\infty} \prod_{l}\prod_{i,j = 1}^{N}
\frac{1}{\pi}
d[{\rm Re}M_{ij}(l)] d[{\rm Im}M_{ij}(l)]\; {\rm e}^{-S_{W}}
\label{zw}
\eeq
where
\beq
S_{W} = \sum_{l} {\rm Tr}[M^{\dagger}(l)M(l)] - \frac{\lambda}{N}
\sum_{\rm plaq} Tr[M(l_{1})M(l_{2})M(l_{3})M(l_{4})]
\label{sw}
\eeq
These two terms in the action are illustrated in figure 1 for $D=2$. `plaq'
indicates a sum over all oriented plaquettes on the lattice. The action has
the usual gauge symmetry associated with unitary transformations at the sites
connected by the links. When one expands the exponential in $\lambda$ and
performs the $M$ integrals one finds the result \cite{wein};
\beq
Z_{W}= \sum_{S} \left(\frac{1}{N}\right)^{\chi} \lambda^{n}
\label{we}
\eeq
where $S$ denotes all closed surfaces made out of plaquettes. $\chi$ is the
Euler number and $n$ the number of plaquettes (the area). The fact that
complex rather than unitary matrices are link variables eliminates the
higher contractions between plaquettes that one finds in the corresponding
expansion of lattice Yang-Mills; the latter gives unwanted contact
interactions between worldsheets. The expansion (\ref{we}) is a
discrete version of the partition
function of 1st quantised Nambu-Goto strings. Evidently this model cannot
be compared with recent results for $c\leq1$ string theory since the plaquettes
require an embedding dimension $D\geq2$.

One can endow the worldsheets of the Weingarten model with an intrinsic metric
by adding the following piece to the action;
\bey
S_{C} &  = & -\sum_{l}  \frac{\kappa_{1}}{N} {\rm Tr}[M^{\dagger}(l)M(l)
M^{\dagger}(l)M(l)] +  \frac{\kappa_{2}}{N}
{\rm Tr}[M^{\dagger}(l)M(l)M(l+1)
M^{\dagger}(l+1)] \nonumber \\ && -  \frac{\kappa_{3}}{N}\sum_{\rm L}
{\rm Tr}[M(l_{1})
M^{\dagger}(l_{1})M(l_{2})M^{\dagger}(l_{2})]
\label{sc}
\eey
whose terms are also illustrated in figure 1. L indicates all L-shaped
terms, as shown. $S_{C}$ thus consists of `collapsed' plaquettes. Note that
orientation is irrelevant for terms contributing to (\ref{sc}); reversing
the link arrows makes no difference. The model with action $S_{W}+S_{C}$ is
remarkably similar to the light-cone lattice string model studied by Klebanov
and Susskind \cite{igor1} (see also \cite{tim1}), although at present the
precise connection is still missing.

If one now expands in $\lambda$ and the $\kappa_{i}$ the effect of the new
terms from $S_{C}$ is to produce insertions on the links of the surfaces of
operators coupling to the $\kappa_{i}$, each insertion giving a factor of
$\kappa_{i}$. One may interpret the latter as the insertion of a surface
element of extrinsic area 0 but intrinsic area 1. The intrinsic and extrinsic
areas of plaquettes in $S_{W}$ are both 1. By allowing collapsed plaquettes
one has generated a dynamical quadrilation (square simplices), representing
an intrinsic metric on the surface \cite{quad}, since a vertex on the
surface may be surrounded by a variable number of plaquettes for given
extrinsic geometry. Put another way, starting from a dynamical quadrilation
one defines the matter field at each vertex to be a height variable
$(h_{1}, \ldots,h_{D})$, where $h_{j}$ is an integer and neighbouring vertices
have $\Delta h_{j} = 0, \pm1$. These height variables are reminiscent of those
of the Solid-On-Solid (SOS) models in statistical mechanics
\cite{baxt}\footnote{For a general introduction see
e.g.\cite{gins}.}, where neighbouring vertices differ in height by 1.
The correspondence can be made
precise by transforming to an auxiliary target lattice. For example if $D=2$,
by drawing the plaquettes on the diagonal lattice of the square
height lattice (figure 2) one finds that $\Delta h_{j} = \pm1$. In higher
dimensions the construction becomes increasingly more complicated. The links
of the auxiliary lattice on which one draws the plaquettes form the body
diagonals of the hypercubes of the height lattice. In other words it is
generated by the (overcomplete) set of vectors $\{(\pm1,\pm1,
\ldots,\pm1)\}$. Note that plaquettes on this lattice need not be planar.
However it is always possible to break them up into triangles so that the
surface element interpretation is not lost. In this way one sees that the
expansion of the new Weingarten model is equivalent to $D$ coupled
SOS models coupled to two-dimensional discrete quantum gravity.
Only in the continuum limit might one expect to see  a decoupling between
the models.

One must now ask for the meaning of the various bare couplings in the model
and what sort of critical behaviour one can expect to obtain. $\frac{1}{N}$
is of course the string coupling. Since every plaquette represents a unit
element of intrinsic area, the parameter $-\log s$ obtained by scaling
$\lambda \rightarrow s\lambda$, $\kappa_{i} \rightarrow s\kappa_{i}$ plays the
r\^ole of worldsheet cosmological constant. Another parameter should set
the target-space scale, the fundamental string tension $\frac{1}{\alpha'}$.
This leaves two couplings in the model $S_{W}+S_{C}$. It is easy to see that
these are not couplings to intrinsic geometry alone since for $D=1$ $\lambda$
and $\kappa_{3}$ do not exist. Neither do they provide coupling to extrinsic
worldsheet curvature (e.g. one cannot even define normals to the plaquettes of
$S_{C}$). This only leaves terms in the worldsheet action consisting of higher
derivatives in the embedding co-ordinate, which are irrelevant. In fact there
are of course an infinite number of naively irrelevant terms one
could put into the
model by including plaquettes of length greater than four. The fact
that they can still affect the continuum limit under approriate
tuning is most clearly illustrated by Kazakov's multicritical points
\cite{kaz}, which will be mentioned in section 2. One is especially
interested in
reproducing Polyakov's formulation of bosonic strings in the
continuum \cite{poly}. In that case the
`link factor' in a dynamical quadrilation is gaussian;
\beq
G(X_{i}-X_{j}) = {\rm exp}\left(-\frac{1}{2}(X_{i}-X_{j})^{2}\right)
\label{pl}
\eeq
where $X_{i}$ and $X_{j}$ are the target space co-ordinates of neighbouring
vertices. For SOS-type matter the link factor is;
\beq
\delta(X_{i} -X_{j} -1) + \delta(X_{i} -X_{j}+1)
\label{wl}
\eeq
In momentum $p$ space (\ref{pl}) is again gaussian ${\rm exp}(-\frac{p^{2}}{
2})$ while the fourier transform of (\ref{wl}) gives $\cos{p}$. These agree
to order $p^{2}$ and so if the critical behaviour is governed by the small
$p$ IR properties one expects to end up in the same universality class
\cite{migd}. This picture could only be upset by UV divergences, in which
case the higher dimension terms in the Weingarten model aquire renewed
importance.

In the next section the continuum limit of the model will be taken for $D=1$,
in which there are no UV divergences. On a flat worldsheet the critical
behaviour of the SOS model is that of one free, massless boson \cite{gins}.
Given all this it will be no suprise to learn that on a fluctuating
worldsheet the critical behaviour is that of a free boson coupled to 2D
gravity, as specified by the hermitian matrix model formulation \cite{gros}.
However the reformulation as a lattice gauge theory has certain conceptual
advantages which will become apparent in section 3.
\section{Continuum Limit for $c\leq1$}
In $D=1$ one retains only the collapsed plaquettes coupling to $\kappa_{1}$
and $\kappa_{2}$, so the action for an $L$-link lattice is;
\bey
S_{D=1} & = & \sum_{l=1}^{L} {\rm Tr}[M^{\dagger}(l)M(l)] -
\frac{\kappa_{1}}{N}
{\rm Tr}[M^{\dagger}(l)M(l)M^{\dagger}(l)M(l)] \nonumber \\ && -
\sum_{l=1}^{L-1}
\frac{\kappa_{2}}{N} {\rm Tr}[M^{\dagger}(l)M(l)M(l+1)M^{\dagger}(l+1)]
\label{S1}
\eey
One can decompose $M(l)$ by a bi-unitary transformation $M(l)=U^{\dagger}(l)
x(l) V(l)$ where $U$, $V$ are unitary matrices and $x$ is a diagonal matrix
with non-negative elements. The measure becomes \cite{tim2};
\beq
{\cal D}U {\cal D}V \prod_{i=1}^{N}dy_{i} \Delta^{2}(y)
\eeq
where $y_{i}=x_{i}^{2}$ is the $i^{\rm th}$ diagonal element, $\Delta$ is the
Van-der-Monde determinant $\prod_{i<j}(y_{i}-y_{j})$ and ${\cal D}U$ is the
Haar measure for the unitary group. Defining $\Omega(l)=U(l)V^{\dagger}(l+1)$
the action (\ref{S1}) becomes;
\beq
\sum_{l=1}^{L}{\rm Tr}\left[y(l) - \frac{\kappa_{1}}{N} y^{2}(l)\right]
- \frac{\kappa_{2}}{N}\sum_{l=1}^{L-1} {\rm Tr}[\Omega^{\dagger}(l)y(l)
\Omega(l)y(l+1)]
\label{S1R}
\eeq
and the measure is;
\beq
\prod_{l=1}^{L} dy(l)\; \Delta^{2}(y(l))\; {\cal D}U(l)\;
{\cal D}V(1) \prod_{l=1}^{L-1} {\cal D}\Omega(l)
\label{M1R}
\eeq
The presence of $L+1$ redundant angular integrals is a reflection of gauge
invariance. The remaining angular integrals can be performed
using the result of Itzykson and
Zuber \cite{itzy} to give the partition function;
\bey
Z_{D=1} & \propto &  \int_{0}^{\infty}\prod_{l=1}^{L}dy(l)
\Delta(y(1))\Delta
(y(L))\;  {\rm exp}\left({\rm Tr}\left[-y(l) + \frac{\kappa_{1}}{N}y^{2}(l)
\right]\right) \nonumber \\
&& \;\;\;\;\;\;\;\prod_{l=1}^{L-1} {\rm det}_{ij} \left[
{\rm exp} \left(\frac{\kappa_{2}}{N}y_{i}(l)
y_{j}(l+1)\right)\right]
\eey
Each determinant is antisymmetric in the $y_{i}$'s and in the $y_{j}$'s but
the whole integrand is symmetric in these variables since $\Delta$ is also
antisymmetric. Thus one can replace each determinant by $N!$ times the first
ordering in its expansion. The only requirement on the range of integration
is that it be non-zero. This is the content of the result of Mehta \cite{mhta};
\bey
Z_{D=1} & \propto & \int_{0}^{\infty} \prod_{l=1}^{L}dy(l)  \Delta(y(1))
\Delta(y(L))\; {\rm exp}( \sum_{l=1}^{L} {\rm Tr} [ -y(l) +
\frac{\kappa_{1}+\kappa_{2}}{N}y^{2}(l)] \nonumber \\
&&\;\;\;\;\;\; - \frac{\kappa_{2}}{2N}
{\rm Tr}[y^{2}(1) + y^{2}(L)] - \sum_{l=1}^{L-1}\frac{\kappa_{2}}{2N} {\rm Tr}
[(y(l+1)-y(l))^{2}] )
\eey
This can be interpreted \cite{migd,gros,bipz} as the discrete-time quantum
mechanical partition function of $N$ non-interacting particles (on $[0,
\infty)$ in the present case) which are in addition fermionic, on account
of the $\Delta$s antisymmetrising the final states at `time' $L$ with respect
to the initial states at `time' 1. For $c=1$, $L \rightarrow \infty$.
The saddle point distribution of
eigenvalues $y_{i}$ that one is interested in is sketched in figure 3. The
non-negativity restriction is denoted by an infinite barrier and the
eigenvalues fill states up to a Fermi level coinciding with the top of the
quadratic maximum by adjusting $\kappa_{1}+\kappa_{2}$. One now proceeds to
the continuum limit by taking a double scaling limit in $N$ and $\kappa_{1}
+ \kappa_{2}$ in the usual way \cite{gros}. Lack of space here prevents a
detailed account and the reader is referred to the review \cite{igor2}
for example. The only novel feature is the presence of the infinite
barrier; but this does not affect the universal critical properties to all
orders in the $\frac{1}{N}$ (genus) expansion at least, since these are
governed purely by the occurrence of a quadratic maximum.

It was shown by Gross and Klebanov \cite{igor3} that in the continuum limit of
the $c=1$ hermitian matrix chain, the transfer matrix between neighbouring
sites in the chain is equivalent to that of the theory on ${\rm I\!R}$, between
those same points, after a redefintion of $\alpha'$ and the string coupling
dependent on the target chain spacing. The same is true here, identifying those
chain sites with the centres of the links of the present model. There is a
range of lattice spacing from
zero to $\pi\sqrt{\alpha'}$ for which one obtains the critical behaviour of
one free boson coupled to 2D gravity. This is the quantum gravity version of
the flat-space fact that, amongst other $c=1$ models, the SOS
model renormalises onto a Gaussian fixed line
on which it has continuously varying critical exponents. This remarkable
short distance property of string theory is nothing more nor less \cite{igor1}.
The transition point of `SOS-strings' has been studied in detail by I.Kostov
in recent times \cite{kost}.

To end this section, consider the finite chain models ($c<1$) \cite{doug}.
The $L$-link linear lattice describes the $L+1^{\rm st}$ restricted
solid-on-solid (RSOS) model coupled to 2D gravity. This target lattice is
precisely that of the $A_{L+1}$ Dynkin diagram (figure 4) and on a flat
worldsheet it is well-known \cite{gins} that these RSOS models at criticality
provide particular realisations of the  unitary discrete series with
$c=1-6/L(L+1)$. Coupling to 2D gravity should again leave them in the same
universality classes as those of the finite hermitian matrix chains,
for reasons
already explained. In particular the $A_{2}$ model;
\beq
S_{A_{2}}= {\rm Tr}[M^{\dagger}M] -\frac{\kappa_{1}}{N} {\rm Tr}[M^{\dagger}
MM^{\dagger}M]
\label{SPG}
\eeq
should describe pure gravity. This has previously been verified in
detail \cite{tim2,jan},
as have Kazakov's multicritical points \cite{kaz,shen,tim4} obtained by
tuning couplings to plaquettes of length $>4$, ${\rm Tr}[M^{\dagger}M]^{p}$.
Lastly, the reader will have noted the omission of discussion on correlation
functions at $c\leq1$. It is easy to see however, from the identification
with hermitian matrix eigenvalues, that the equivalence of correlators
punctual in target
space carries over also since these involve only eigenvalues. A proof that
appropriately defined string states on the lattice can yield the standard
correlators
on a continuous target space, in any context, is still lacking.
\section{Continuum limit on $S^{1}$}
In this section the continuum limit of the model defined on a periodic
target lattice will be explored. One may think of these lattices as the
extended Dynkin diagrams $\hat{A}_{L-1}$ (figure 4). Recalling the expression
(\ref{S1R}),  there will now be a coupling;
\beq
-\frac{\kappa_{2}}{N} {\rm Tr}[\Omega^{\dagger}(L) y(L) \Omega(L) y(1)]
\eeq
where $\Omega(L) = U(L)V^{\dagger}(1)$. However the measure (\ref{M1R}),
 unlike  that of the hermitian
matrix model, allows one to again perform all angular integrals using Itzykson
and Zuber's formula to derive;
\beq
Z_{S^{1}} = {\rm const} \int_{0}^{\infty} \prod_{l=1}^{L}dy(l) {\rm exp}
\left( {\rm Tr}\left[ -y(l) + \frac{\kappa_{1}}{N} y^{2}(l)\right]\right)
{\rm det}_{ij} \left[ {\rm exp}
\left(\frac{\kappa_{2}}{N}y_{i}(l)y_{j}(l+1)\right)
\right]
\eeq
where $y(L+1) \equiv y(1)$. This represents $N$ non-interacting
non-relativistic fermions on $[0,\infty)$ at finite temperature. The angular
degrees of freedom could be eliminated because the action (\ref{S1})
explicitly excludes vortex configurations, which are the plaquettes which
wind around $S^{1}$, ${\rm Tr}[\prod_{l=1}^{L} M(l)]^{p}$. These statements
will now be justified in more detail.

In order to show equivalence to non-interacting fermions it is convenient to
take the simplest case of  $L=1$ i.e. a single periodic link.
The argument generalises easily to $L>1$ and could equally have been
used in the previous section. Note that the
target lattice $\hat{A}_{0}$ (which is not a Dynkin diagram) is usually
omitted from the $\hat{A}\hat{D}\hat{E}$ classification \cite{gins}, and
has no analogue in the hermitian matrix models.
Periodic $L=1$ is the complex matrix model;
\bey
Z_{\hat{A}_{0}} & = & \int dM\; {\rm exp}(
 -{\rm Tr}[M^{\dagger}M]
+ \frac{\kappa_{1}}{N}{\rm Tr}[M^{\dagger}MM^{\dagger}M] +\frac{\kappa_{2}}{N}
{\rm Tr}[M^{\dagger}MMM^{\dagger}])  \label{ZV}\\
& = & {\rm const}\int_{0}^{\infty} \prod_{k=1}^{N} dy_{k}\; {\rm exp}
\left(-y_{k} +\frac{\kappa_{1}}{N}y^{2}_{k} \right)
{\rm det}_{ij} \left[ {\rm exp} \left( \frac{\kappa_{2}}{N}y_{i}y_{j}\right)
\right]
\label{ZV2}
\eey
The restriction to a periodic gauge invariance $M \rightarrow U^{\dagger}
MU$ has meant that one cannot transform to eigenvalues directly as happens for
(\ref{SPG}). As an aside, one notes that $Z_{\hat{A}_{0}}$ describes the vertex
model of statistical mechanics on a dynamical quadrilation. One might expect
this from the SOS $\leftrightarrow$ vertex correspondence.
More directly, if one expands $Z_{\hat{A}_{0}}$ in Feynman diagrams using
double-line notation, the propagators carry an orientation since the matrices
are complex (figure 5). In fact these are the dual diagrams to the those of
the plaquettes on the worldsheet. The notion of orientation of one vertex
with repect to its neighbour is significant in defining the regular lattice
vertex model, but this distinction is lost when coupling to
gravity. For this reason the random lattice version (\ref{ZV}) is the analogue
of a restricted type of vertex model known as the ferro-electric or F-model
\cite{gins}.

The clearest way to show that (\ref{ZV2}) represents non-interacting
non-relativistic fermions at finite temperature is to introduce a complete
set of antisymmetric wavefunctions\footnote{I thank
I.Klebanov for suggesting to
argue in this way.}. In detail, if $\{ \psi_{\alpha}\}$ are a complete set
of one-body wavefunctions (the precise hamiltonian being unimportant initially)
then $\sum_{\alpha}\psi_{\alpha}^{\dagger}(y)\psi_{\alpha}(z)=\delta(y-z)$
and for any function $S(y)\equiv \prod_{i} S(y_{i})$;
\bey
&&\int \prod_{k} dy_{k} {\rm det}_{ij}\left[ {\rm exp} \left(
\frac{\kappa_{2}}{N}y_{i}y_{j} \right) \right] S(y) \nonumber \\
& = & \sum_{\Pi(j_{1},\ldots,j_{N})} \epsilon_{j_{1},\ldots,j_{N}}
\int\prod_{k} dy_{k}dz_{k} \sqrt{S(y)} \sqrt{S(z)}{\rm exp}\left(
\frac{\kappa_{2}}{N}y_{k}z_{k}\right) \delta(y_{k} -z_{j_{k}})
\label{mess}
\eey
Rewriting the $\delta$ functions using the fact that;
\bey
&&\sum_{\Pi} \epsilon_{j_{1},\ldots,j_{N}} \prod_{k=1}^{N} \sum_{\alpha_{k}}
\psi_{\alpha_{k}}^{\dagger} (y_{k}) \psi_{\alpha_{k}}(z_{j_{k}}) \nonumber \\
& = &
\frac{1}{N!} \sum_{\alpha_{1},\ldots,\alpha_{N}} \left(\sum_{\Pi}
\epsilon_{\alpha_{1},\ldots,\alpha_{N}}\prod_{k} \psi_{\alpha_{k}}^{\dagger}
(y_{k})\right)\left(\sum_{\Pi} \epsilon_{\alpha_{1},\ldots,\alpha_{N}}
\prod_{k} \psi_{\alpha_{k}} (z_{k})\right) \\ & \equiv & \sum_{A}<\!
\Psi_{A}|y\!><\!z| \Psi_{A}\!>
\eey
where $\Psi_{A}$ are a complete set of Slater determinants (by definition
essentially), (\ref{mess}) becomes;
\bey
&&\int dy dz \sum_{A} <\! \Psi_{A}|y\!>\sqrt{S(y)}\; {\rm exp}\left(
\frac{\kappa_{2}}{N} \sum_{k} y_{k} z_{k}\right) \sqrt{S(z)} <\! z| \Psi_{A}
\!> \\
&=& \int dy dz \sum_{A} <\! \Psi_{A}|y\!><\!y|{\rm e}^{-\beta H}|z\!><\!z|
\Psi_{A}\!> \;\equiv {\rm Tr}_{A} {\rm e}^{-\beta H}
\eey
for an appropriate sum of one-body hamiltonians $H=\sum_{k} h_{k}$. The
transfer matrix elements for each $h_{k}$ in the present case are;
\beq
{\rm exp}\left( -\frac{\kappa_{2}}{2N}(y_{k}-z_{k})^{2} - \frac{1}{2}
\left( y_{k}-\frac{\kappa_{1}+\kappa_{2}}{N}y_{k}^{2}\right)
-\frac{1}{2}\left( z_{k} - \frac{\kappa_{1}+\kappa_{2}}{N}z_{k}^{2}
\right)\right)
\label{TM}
\eeq
The analysis now proceeds much as in ref.\cite{igor3}. (\ref{TM}) is a
particular example of a  class of transfer matrices which are equivalent,
in the double scaling limit, to that of the harmonic oscillator hamiltonian.
The only requirement is that $V(y) = y -\frac{\kappa_{1} + \kappa_{2}}{N}y^{2}$
(figure 3) possess a quadratic maximum. Thus $Z_{\hat{A}_{0}}$ is the
partition function of fermions at temperature $1/\epsilon$, where $\epsilon$
is the length of a single link. The equivalence with the harmonic
oscillator breaks down for $\epsilon >\pi\sqrt{\alpha'}$ \cite{igor3} when
there is presumably a momentum-mode condensation in the same manner as that of
the regular lattice vertex model\footnote{The temperature seems to be out by
a factor of 2 which the author does not fully understand. A similar factor
of 2 arises in the Klebanov-Susskind model due to a mismatch
of Hilbert spaces \cite{tim1}}.
Generalising the previous analysis to $\hat{A}_{L-1}$ shows that this
describes a compact boson with radius $0<r<L\sqrt{\alpha'}/2$ coupled to
2D gravity in accordance with the hermitian matrix models. On a circle of
circumference $\epsilon L$ there are Kaluza-Klein momentum modes $\cos{(2\pi
nX/\epsilon L)}$ of conformal weight $\pi^{2}n^{2}\alpha'/\epsilon^{2}
L^{2}$. For $c=1$ the conformal weight $\Delta_{0}$ of an operator and its
gravitational scaling dimension $\Delta$ are simply related, $\Delta^{2} =
\Delta_{0}$. Given that $\gamma_{\rm str} =0$, this means that coupling to
worldsheet gravity should not change the radius at which operators become
relevant.  Thus it would seem that the
lowest dimension operator in the action for the model on an $L$-link
periodic lattice
corresponds to momentum $n=L$. At the critical radius this corresponds to a
rather trivial example of perturbation by one of the so-called special
operators of $c=1$ string theory.

The dual operators which also govern the phase structure are the vortices
(winding mode condensates). These are given by the Wilson line operators for
(anti) vortex number $(q) p$;
\beq
L_{p} = {\rm Tr}\left[ \prod_{l=1}^{L} M(l) \right]^{p} \;,\;
L_{-q} = {\rm Tr} \left[ \prod_{l=1}^{L} M^{\dagger}(l)\right]^{q}
\label{LL}
\eeq
which were explicitly omitted from the action up to now.
Although their inclusion leaves
one with angular degrees of freedom which cannot be eliminated with standard
methods, their expression in the familiar form (\ref{LL})
represents a considerable
simplification over the hermitian matrix formulation which organises the
contribution of vortices unusually.
For example the simplest two-point function, which can be evaluated in the
$Z_{\hat{A}_{0}}$ theory without loss of generality, involves the
angular integral;
\beq
\int d\Omega \; {\rm Tr}[\Omega \sqrt{y}] {\rm Tr}[\Omega^{\dagger} \sqrt{y}]
\; {\rm exp}(\Omega y \Omega^{\dagger} y)
\eeq
which arises in models of unitary matrices in external fields.
Since one expects
duality to be a good symmetry in the $1/N$ expansion even after
coupling to worldsheet gravity \cite{igor3}, the vortex operators should
become relevant at the appropriate dual radii. This locates the first
possible transition at $r=2\sqrt{\alpha'}$ \cite{igor4} where
$<\!L_{1}L_{-1}\!>$ will
scale correctly.
\section{Discussion}
To summarise, the main result of this letter has been to rewrite
 non-critical bosonic string theory in the form of a lattice
gauge theory.
There are many
infamous parallels between the confining phase of Yang-Mills and string
theory. In particular the finite temperature deconfining transition of
pure Yang-Mills with $U(1)$ group centre\footnote{For an
introduction to aspects of deconfinement in (lattice) gauge theories see
e.g. \cite{svet}.} is analogous to a phase transition conjectured to lie
at (or near) the Hadgedorn temperature $T_{H}$ of critical string theory
\cite{Sath}, interpreted as a condensation of winding modes.
It is natural to ask what the equivalence proved in this letter can tell
one about the nature of this transition in general, since in higher
dimensional string theories one hopes to expose the underlying degrees
of freedom in the high-temperature phase \cite{witt}. In Yang-Mills theory
this further tier of simplification reveals the glue.

It is
the presence of a global $U(1)$ symmetry of the local action, whose charge
is winding number, which ties together seemingly disparate models in the
picture of deconfinement \cite{svet}. Under $M \rightarrow
{\rm e}^{i\theta}M$, for link matrices around a compact direction, the Wilson
 lines are
not invariant, $L_{p} \rightarrow
{\rm e}^{ipL\theta}L_{p}$. $<L_{1}>$ is then an order
parameter for the transition
to a phase in which the vacuum breaks this $U(1)$ symmetry spontaneously.
This picture is only strictly correct in target spaces of dimension
$>2$ due to usual the prohibition of spontaneous breakdown
of continuous symmetry. In the case of section 4, the effective
action for the field $L_{1}$, defined by;
\beq
{\rm e}^{-S[L_{1}]} = \int dM \delta\left(L_{1}-{\rm Tr}\left[
\prod_{i=1}^{L} M(l_{i})\right]\right) \;{\rm e}^{-S_{W}-S_{C}}
\eeq
is zero dimensional. $S[L_{1}]$ is symmetric under $U(1)$ and it alone
is a reliable guide to the phase structure.
The transition is then of Kosterlitz-Thouless-Berezinski
type \cite{igor4}, where $<\! |L_{1}|\!>$ is the order parameter and $U(1)$
winding symmetry remains unbroken. From the worldsheet point of view the
transition would always be of BKT type whatever the target space dimension.
Simply speaking one cannot span a winding loop with a surface. Thus there
is no particularly fundamental difference between the target space and
worldsheet descriptions in low dimensions. One expects the high temperature
phase to also be a random surface theory. In higher dimensions it is possible
to break $U(1)$ and have a true deconfining transition, where $<\! L_{1} \!>
\sim {\rm e}^{-N}$. Unfortunately it is precisely in these higher dimensions
that the simplest interacting bosonic theory does not present one with a
conventional random surface representation at zero temperature. Nevertheless
it may be fruitful (if a little naive) to re-examine the one-dimensional
Weingarten model as a more or less conventional gauge theory. While
the effectively 2 field theoretic degrees of freedom argued
to underlie the $c=26$
critical string \cite{witt} are realised trivially in the
$c=1$ non-critical string,
such an examination may  shed light on a non-perturbative framework
for string theory.
\vspace{10mm}

\noindent {\bf Acknowledgements:} I am indebted to Tim Morris for
arousing my (belated) interest in this problem and for generously sharing
with me his results \cite{timx}.
I also thank Mike Newman and especially
Igor Klebanov for valuable conversations. This work was supported by
S.E.R.C.(U.K.) post-doctoral fellowship RFO/B/91/9033.
\vfill
\newpage
\begin{center}
{\bf Figure Captions}\\
\end{center}
\begin{flushleft}
Figure 1 : Terms in $S_{W}$ (the propagator and $\lambda$-plaquette) and
$S_{C}$ (the $\kappa_{1},\kappa_{2},\kappa_{3}$-collapsed plaquettes).\\
\vspace{5mm}
Figure 2 : The auxiliary lattice for $D=2$, such that $\Delta h_{1,2} = \pm 1$.
Only terms from $S_{W}$ are shown.\\
\vspace{5mm}
Figure 3 : The single-eigenvalue potential for $S_{D=1}$ near criticality.\\
\vspace{5mm}
Figure 4 : The (extended) Dynkin diagram target lattices.\\
\vspace{5mm}
Figure 5 : Feynman diagram vertices for $Z_{\hat{A}_{0}}$. The propagator is
$<\!M_{ab}M_{cd}^{\dagger}\!> = \delta_{ad}\delta_{bc}$.
\end{flushleft}
\vfil

\vfil
\end{document}